\documentclass[aps,epsf,epsfig,twocolumn]{revtex4}
\usepackage[dvips]{graphicx}
\usepackage[usenames]{color}
\usepackage{amssymb}
\usepackage{amsmath}

\def\be{\begin{equation}}
\def\ee{\end{equation}}
\def\bea{\begin{eqnarray}}
\def\eea{\end{eqnarray}}
\def\bi{\begin{itemize}}
\def\ei{\end{itemize}}
\def\bin{\begin{enumerate}}
\def\ein{\end{enumerate}}

\begin{document}

\title{Quantum phase transition and critical fluctuations of an attractive Bose gas in a double well potential}

\author{P. Zi\'n,$^1$ J. Chwede\'nczuk,$^2$ B. Ole\'s,$^3$ K. Sacha$^3$ and M. Trippenbach,$^{1,2}$}

\affiliation{$^1$Soltan Institute for Nuclear Studies, Ho\.za 69,
00-681 Warsaw, Poland}

\affiliation{$^2$Institute for Theoretical Physics, Warsaw University,
Ho\.za 69, 00-681 Warsaw, Poland}

\affiliation{$^3$Marian Smoluchowski Institute of Physics and
Mark Kac Complex Systems Research Centre, Jagiellonian University,
Reymonta 4, 30-059 Krak\'ow, Poland}

\begin{abstract}
We consider a Bose gas with an attractive interaction in a symmetric double well potential. In the Hartree approximation, the ground state solution spontaneously breaks the symmetry of the trapping potential above certain value of the interaction strength. We demonstrate how the Landau-Ginzburg scheme of the second order phase transition emerges from the quantum model and show its link to the spontaneous symmetry breaking mentioned above. We identify the order parameter, the critical point and analyze quantum fluctuations around it.
\end{abstract}

\maketitle

%%%%%%%%%%%%%%%%%%%%%%%%%%%%%%%%%%%%%%%%%%%%%%%%%%%%%%%%%%%%%%%%%%%%%%%%%%%%
%\section{Introduction}

Ultra cold atomic gasses become ideal systems to investigate many body physics both theoretically and experimentally \cite{Keter}. Experimentalists can prepare quantum degenerate gasses in a wide range of trapping potentials and tune strength and even character of particle interactions \cite{Feshbach}. Various phase transitions in ultra-cold Bose gases have been already considered \cite{Bloch,Esslinger,ueda}. For example it is shown in Ref.~\cite{ueda} that a finite-size, translationally invariant, one-dimensional Bose gas with attractive interactions above critical value is fragile against a weak symmetry-breaking perturbation. The authors report a crossover from a translationally symmetric ground state to localized state.

In this Letter we study a Bose gas with attractive interaction trapped in a symmetric double well potential. To the first approximation the ground state of such many body system is analyzed using Hartree approach, where it is assumed that all the atoms are in the same single-particle state (mode). This treatment leads to the non-linear Schr\"odinger equation, known as Gross-Pitaevskii Equation (GPE) \cite{GPEcite}. The GPE is extremely successful in the analysis of systems of ultra-cold bosons and the single particle mode is called the ``condensate mode''. When this equation is applied to the double well system, it turns out that the ground state solution of the GPE breaks the reflection symmetry of the double well problem, when the interaction strength exceeds some critical value. A better insight into the properties of the system is gained within the Bogoliubov method \cite{Leggett}. In contrast to the GPE, it does not neglect the population of the non-condensate modes, although their treatment is approximate. When the interaction strength in the double well system approaches a critical value, the Bogoliubov theory reveals gapless spectrum, diverging number of particles in the non-condensate modes and simply breaks down. Hence in the vicinity of the critical value one has to analyze the properties of the ground state solution of the full quantum Hamiltonian.

In present study we use tight binding approximation and apply the Bose-Hubbard model \cite{Czyrak,Leggett,Spekkens,Gati}. We show, using a simple approximation (called a continuum approximation \cite{Spekkens,Juha}) that the Schr\"odinger equation obtained from our Bose-Hubbard Hamiltonian reduces to a one-dimensional Schr\"odinger-like equation of a fictitious particle in an effective potential. Evolution of the shape of this potential from a parabolic, through quartic, to double well as we change the interaction strength reflects second order quantum phase transition. Finally we analyze quantum fluctuations in vicinity of the critical point.

%Finally, we intended to mention several very closely related studies \cite{Spekkens,Corney,cpt,Konotop,Cirac1,Juha}, but we feel the more appropriate place to do so is after we introduce our approach. So we postpone our comments on that and present it below.

%%%%%%%%%%%%%%%%%%%%%%%%%%%%%%%%%%%%%%%%%%%%%%%%%%%%%%%%%%%%%%%%%%%%%%%%%%%%
%\section{Mean field description}
%\label{mean}

The Hamiltonian of atoms trapped in a double well potential reads
\begin{equation}\label{Ham1}
    \hat{H} = -\frac{\Omega}{2}\left( \hat{a}_1^{\dagger} \hat{a}_2
        + \hat{a}_2^{\dagger} \hat{a}_1 \right) + \frac{U}{2}
        \left( \hat{a}_1^{\dagger}\hat{a}_1^{\dagger} \hat{a}_1 \hat{a}_1
        + \hat{a}_2^{\dagger}\hat{a}_2^{\dagger} \hat{a}_2 \hat{a}_2 \right),
\end{equation}
where $\hat a_1$ ($\hat a_2$) operator annihilates an atom in the left (right) well, $\Omega$ stands for the tunneling rate between the wells and $U$ is the on-site interaction strength. Here we consider attractive interactions hence $U$ is negative. Total number of particles $N$ remains a conserved
quantity. We can extract a constant part from the Hamiltonian (\ref{Ham1})
\begin{eqnarray}
\hat{H} & = & -\frac{\Omega}{2}\left( \hat{a}_1^{\dagger} \hat{a}_2
+  \hat{a}_2^{\dagger} \hat{a}_1 \right) + \frac{U}{4} \left( \hat a_1^\dagger \hat a_1 -\hat a_2^\dagger \hat a_2 \right)^2 \nonumber \\
& + & \frac{U}{4}\left(\hat{N}^2 - 2\hat{N}\right),\label{Ham2}
\end{eqnarray}
which will be neglected in the further considerations.

Let us review here some properties of the Hartree approximation applied to the two-mode problem.
It assumes all the atoms being in the same quantum state. The most general form of such a state is
\begin{equation}
\frac{1}{\sqrt{N!}} \left( \sqrt{\frac{1+z}{2}}e^{i\varphi/2} \hat a_1^\dagger + \sqrt{\frac{1-z}{2}}e^{-i\varphi/2} \hat a_2^\dagger \right)^N |\mbox{vac} \rangle. \label{clasic}
\end{equation}
Here,  $z$ is the relative population difference between the wells, $\varphi$ is a relative phase and $|\mbox{vac} \rangle $ denotes the vacuum state. The Hartree method relies on calculating an expectation value of the Hamiltonian (\ref{Ham2}) in the state (\ref{clasic}), which gives
\be\label{newhamil}
\langle \hat H \rangle =\frac{\Omega
N}{2}\left(\frac{\gamma}{2}z^2-\sqrt{1-z^2}\cos\varphi\right).
\ee
The dimensionless parameter
\be
\gamma={UN}/{\Omega},
\ee
is a ratio of the on-site interaction per atom to the tunneling rate. Notice that the minimum of the expectation value of the Hamiltonian (\ref{newhamil}) occurs for $\varphi=0$. In case of $\gamma\geq -1$ the minimum appears for $z = 0$ and for $\gamma < -1$ it is shifted to the point $\pm z_0$, where $z_0=\sqrt{1-\gamma^{-2}}$. The non-zero value of $z_0$ means, that solution of the Hartree approximation breaks the symmetry of the trapping potential. This suggests that a second order phase transition takes place at $\gamma=-1$ in the corresponding quantum model. It is a main purpose of our investigation to confirm this conjecture.

%In conclusion the mean field description (in which all the atoms occupy the same state) reveals that when $\gamma$ decreases and passes through $-1$ there is a change of the ground state from a symmetric superposition of the modes localized in the wells,
%\be
%\phi_0=\frac{1}{\sqrt{2}}\left(\psi_1+\psi_2\right),
%\label{phi0s}
%\ee
%to the case where they occupy one of the asymmetric wavefunctions
%\be
%\phi_{0,\pm}=\sqrt{\frac{1\pm z_0}{2}}\psi_1+\sqrt{\frac{1\mp z_0}{2}}\psi_2.
%\label{phi0a}
%\ee

%%%%%%%%%%%%%%%%%%%%%%%%%%%%%%%%%%%%%%%%%%%%%%%%%%%%%%%%%%%%%%%%%%%%%%%%%%%
%\section{Continuous description}
%\label{contsec}
%
%To show it we introduce below an approach known as the continuum approximation, which was derived in \cite{Spekkens,Juha} for the case repulsive interaction.
%Below we shortly present the derivation and analyze the quality of this approximation in the attractive interaction case.
%that allows us to analyze the double-well problem in a wide range of the system parameters and obtain better insight into physical properties around the critical point.

As the total number of atoms is conserved, the wave function can be written in a Fock basis of states
$|N-n,n \rangle$ (here $n$ denotes number of atoms in the right potential well), as
\begin{equation}
|\Psi \rangle = \sum_{n=0}^N \psi_n |N-n,n \rangle.
\label{fockb}
\end{equation}
Then the stationary Schr\"odinger equation obtained from
Hamiltonian (\ref{Ham2}) reads
\begin{equation*}
\sum_{n}\left(\frac{\Omega}{2}\left[\psi_{n+1}\sqrt{(N-n)(n+1)} + \psi_{n-1} \sqrt{(N-n+1)n}\right]
\right.
\end{equation*}
\begin{equation}
\left.- \frac{U}{4} \psi_n (N-2n)^2 + E \psi_n  \right) |N-n,n \rangle = 0.\label{cont1}
\end{equation}
Let $z_n$ denote the variable related to the relative population difference,
\be
z_n = \frac{(N-n)-n}{N}=1 - \frac{2n}{N}.
\ee
Using this variable, Eq.(\ref{cont1}) transforms into
\begin{eqnarray}\label{recur}
&&\frac{\Omega N}{2}\left[-\psi_{n+1}\sqrt{\frac{1+z_n}{2}\left( \frac{1-z_n}{2} + \frac{1}{N} \right)}+\right.\\
&&\left.\psi_{n-1} \sqrt{ \left( \frac{1+z_n}{2} + \frac{1}{N} \right)\frac{1-z_n}{2}}\right] +
\frac{UN^2}{4} \psi_n z_n^2 = E\psi_n \nonumber.
\end{eqnarray}
For a large number of particles the ``$\frac{1}{N}$'' terms  can be dropped \cite{remark1}. Also in this limit the variable $z_n$ may be treated as a continuous variable (i.e. $z_n \rightarrow z$) and we approximate the finite differences by a second order differential operation
\begin{equation}
\frac{ \psi_{n+1} +\psi_{n-1} - 2\psi_n}{(2/N)^2} \simeq \frac{\mbox{d}^2}{\mbox{d} z^2} \psi(z).
\end{equation}
Upon completing this procedure we reduce Eq.(\ref{recur}) to the following form
\begin{eqnarray}\label{glowne}
\frac{\Omega N}{2}\left(-\frac{2}{N^2}\sqrt{1-z^2}\frac{\mbox{d}^2}{\mbox{d} z^2}+V(z)\right) \psi(z)   =
E \psi(z),
\end{eqnarray}
where we introduced the effective dimensionless potential $V(z)$ 
\be
V(z)=-\sqrt{1-z^2}+\frac{\gamma}{2} z^2,
\label{veff}
\ee
with $\gamma$ defined above. Notice that operator on the left hand side of Eq.(\ref{glowne}) is non-hermitian. However, our starting Eq.(\ref{recur}) is a result of an action of a hermitian operator (\ref{Ham2}) on the wave-function (\ref{fockb}).
In principle, we could obtain the continuous limit in way conserving hermiticity, but the derivation would require much more sophisticated expansion and would loose its illustrative character. Instead we present an alternative derivation and we argue that the corrections required to restore hermiticity in Eq.(\ref{glowne}) are negligible in the large $N$ limit.

An alternative derivation employs the concept of relative phase operator~\cite{Pitaevskii}.
%To obtain the hermitian form we need to discuss another possible derivation of the continuum approximation.
This operator in general is ill-defined, but still is believed to give correct results in the large $N$ limit. We start from the classical Hamiltonian 
\begin{eqnarray}
H = \frac{\Omega N}{2} \left(\frac{1}{2}\sqrt{1- z^2} (1-\cos  \varphi) + \right.
\nonumber \\
\left. +\frac{1}{2}(1-\cos \varphi)\sqrt{1- z^2}   + V(z)  \right),
\label{htakie}
\end{eqnarray}
where $V(z)$ is defined by Eq.(\ref{veff}) (see for example \cite{Drummond}). Next we perform canonical quantization by changing the conjugate variables $z$ and $\varphi$ into operators with the commutation relation
$[\hat \varphi,\hat z] = \frac{2i}{N}$ \cite{Pitaevskii,Leggett}. In the case of repulsive interactions, the phase fluctuations are dominant over fluctuations of population imbalance $z$. Thus in this case  it is convenient to choose phase representation, i.e. replace $\hat z$ with $\frac{2i}{N} \frac{\mbox{d}}{\mbox{d} \varphi} $ \cite{Pitaevskii}.
%It is worth noticing that in this representation the notion of the relative phase operator can be made exact \cite{Drummond}.
On the other hand, in the case of attractive interactions the phase fluctuations are small $\cos\hat \varphi\approx 1-\frac12 \hat\varphi^2$ and it is more convenient to choose $z$ representation;  $\hat \varphi = \frac{2i}{N} \frac{\mbox{d}}{\mbox{d} z}$. In this case we obtain
\begin{equation}
\hat H = \frac{\Omega N}{2} \left( -\frac{1}{N^2} \sqrt{1- z^2}\frac{\mbox{d}^2}{\mbox{d} z^2}
- \frac{1}{N^2}\frac{\mbox{d}^2}{\mbox{d} z^2} \sqrt{1- z^2}  + V(z)  \right).
\label{htakie1}
\end{equation}
The Schr\"odinger equation takes the form
\begin{eqnarray}\nonumber
&& -\frac{1}{N^2} \left(\sqrt{1- z^2}\frac{\mbox{d}^2}{\mbox{d} z^2}  + \frac{\mbox{d}^2}{\mbox{d} z^2} \sqrt{1- z^2}\right)\psi(z) +
\\
&& + V(z)\psi(z) =  \frac{2E}{N\Omega}\psi(z).\label{finaleq}
\end{eqnarray}
This is again a one dimensional Schr\"odinger-like equation for a particle in the potential $V(z)$, where the coefficient $2/N$ plays the role of $\hbar$. One can compare Eq.(\ref{glowne}) and Eq.(\ref{finaleq}) and notice the the difference consists of two terms
\begin{equation*}
\frac{1}{N^2}\frac{z}{\sqrt{1-z^2}}\frac{\mbox{d}}{\mbox{d} z}\psi(z) -
\frac{1}{N^2} \left(\frac{\mbox{d}^2}{\mbox{d} z^2}\sqrt{1-z^2} \right)\psi(z).
\end{equation*}
To estimate the first term we notice that the characteristic width of the wavefunction $\psi$ scales like certain positive power of $1/N$ (since $2/N$ plays the role of $\hbar$). Hence each derivative over $z$ generates a factor of the order of positive power of $N$, and so the first derivative is much smaller than the second one. The second term contributes to the potential and contains a factor of $1/N^2$.  In conclusion both these terms can be neglected.

\begin{figure}[tbp]
\includegraphics[width=9cm]{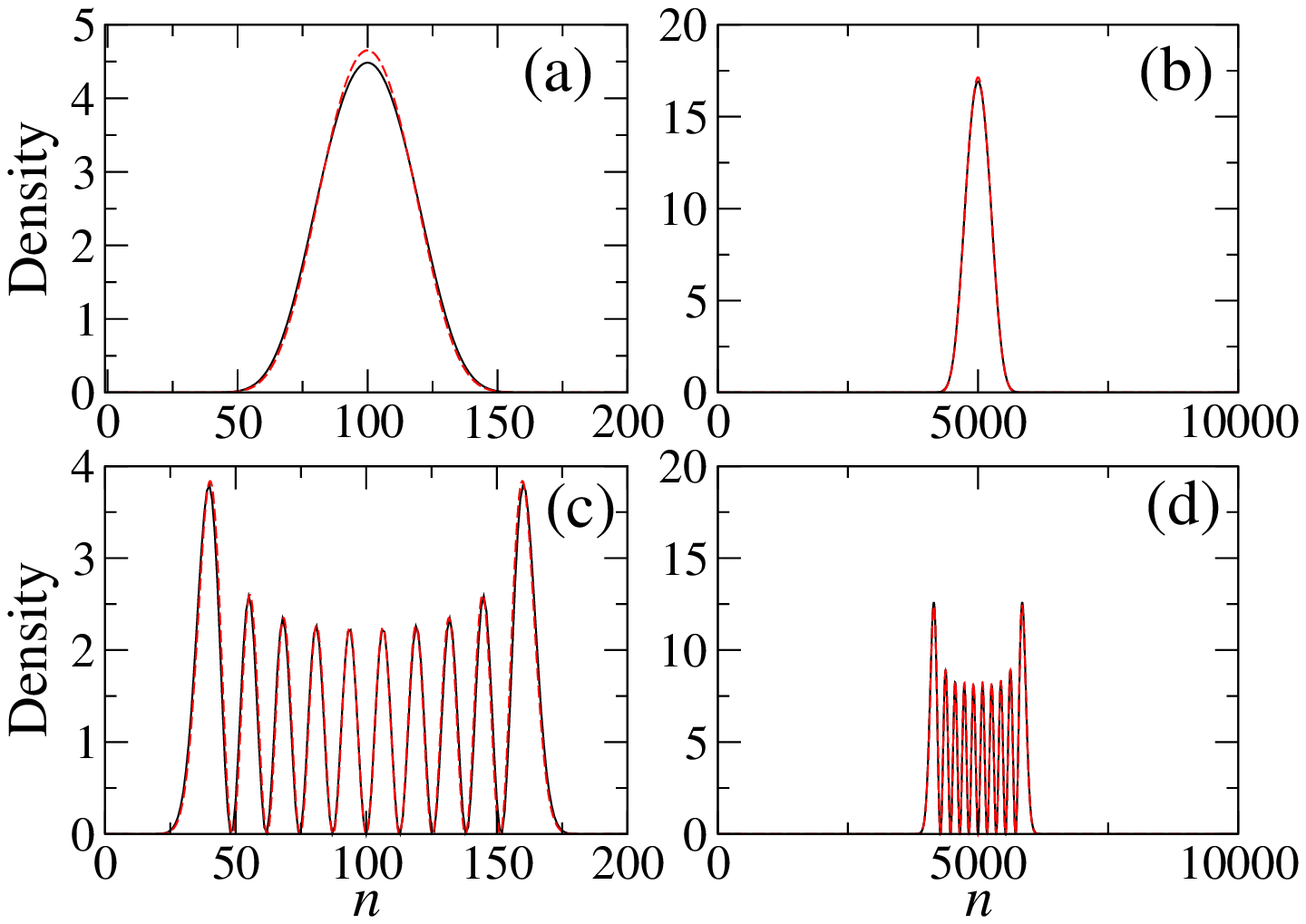}
\caption{(color online). Comparison of the probability density $|\psi(z_n)|^2$ obtained by diagonalization
of the exact Hamiltonian (\ref{Ham2}) (solid black lines) and by solution of Eq.(\ref{finaleq}) (dashed red lines) at the critical point, i.e. for $\gamma=-1$. Panels~(a) and (b) correspond to ground states while (c) and (d) to the ninth excited states. Results for $N=200$ are presented in (a) and (c) and for $N=10^4$ in (b) and (d).}
\label{fig0}
\end{figure}

\begin{figure}[tbp]
\includegraphics[width=8.5cm]{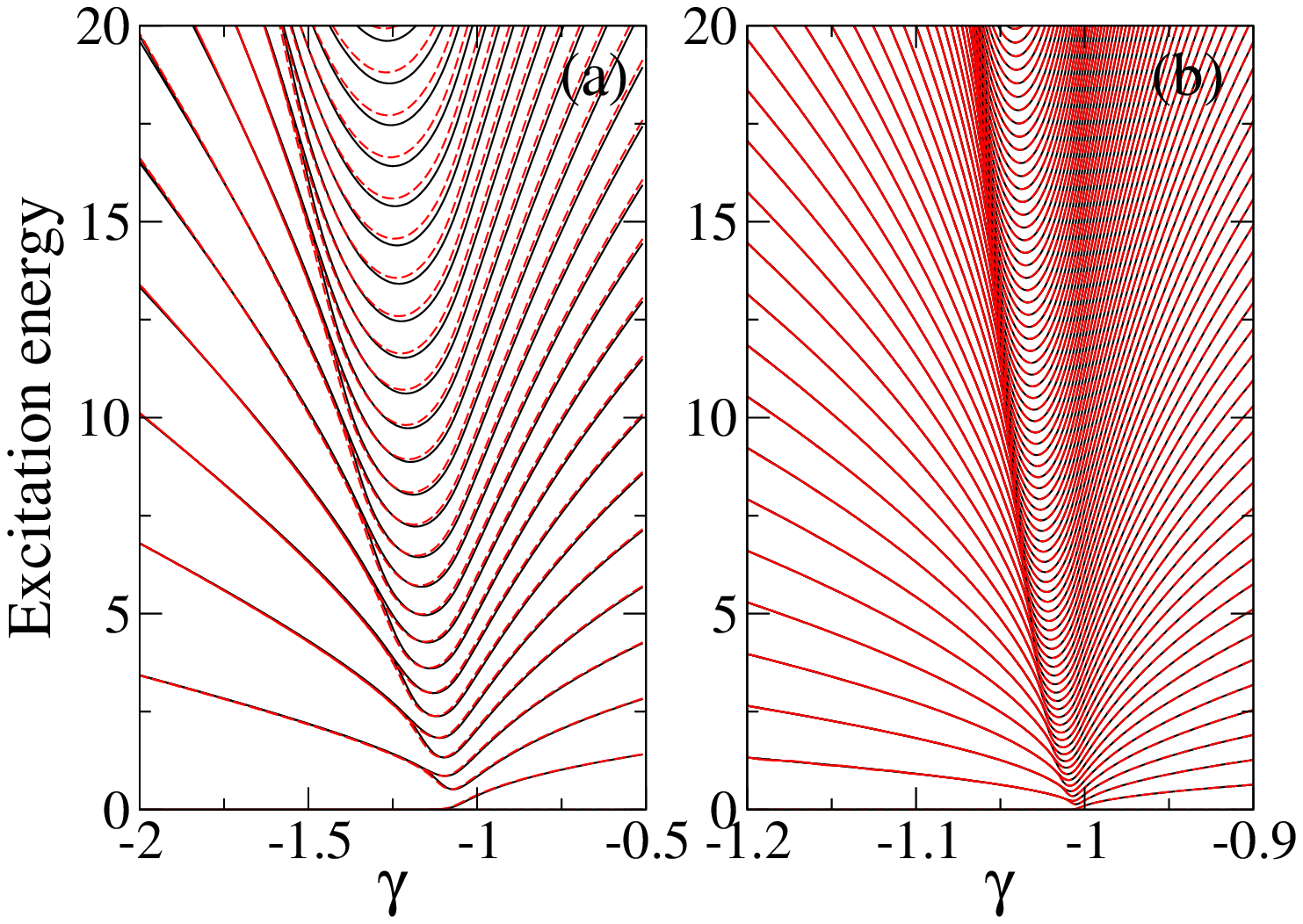}
\caption{(color online). Comparison of the energy levels (i.e. $2(E-E_0)/\Omega$ where
$E_0$ is the ground state energy) obtained by diagonalization of the exact Hamiltonian (\ref{Ham2}) (solid black lines) and by solution of Eq.(\ref{finaleq}) (dashed red lines) versus $\gamma$ for $N=200$ (a) and $N=10^4$ (b).}
\label{fig1}
\end{figure}

Note that Eq.(\ref{finaleq}) turns out to generate accurate results even for relatively low number of atoms. In order to evaluate the quality of the continuum approximation we compare the eigenstates (Fig.\ref{fig0}) and the energy level diagrams (Fig.\ref{fig1}) obtained by solving Eq.(\ref{finaleq}) and by numerical diagonalization of the Hamiltonian (\ref{Ham2}) for $200$ and $10^4$ particles. Already at the level of few hundred particles, the results obtained within the continuum approximation seems to be indiscernible from the exact ones.

Another interesting issue is the relation between the expectation value of the full many body Hamiltonian in the Hartree approach and our continuum approximation. The potential $V(z)$ in Eq.(\ref{veff}) is {\it exactly}  the same as an expectation value of the Hamiltonian in Eq.(\ref{newhamil}), provided we set $\phi=0$. The relation between them is established by the quantization procedure presented above. Recall that at the starting point of this procedure there was a classical Hamiltonian, which turns out to have exactly the same functional dependence as the expectation value of many body Hamiltonian in the Hartree approximation (compare Eq.(\ref{newhamil}) and Eq.(\ref{htakie})). In the quantization procedure the terms with $(1-\cos  \varphi)$ transform into the kinetic energy like operators, leaving $V(z)$ as an effective potential. Hence the same potential is responsible for spontaneous symmetry breaking in the Hartree approach and for the quantum phase transition described by the many body quantum theory.

\begin{figure}
\includegraphics[width=7.5cm]{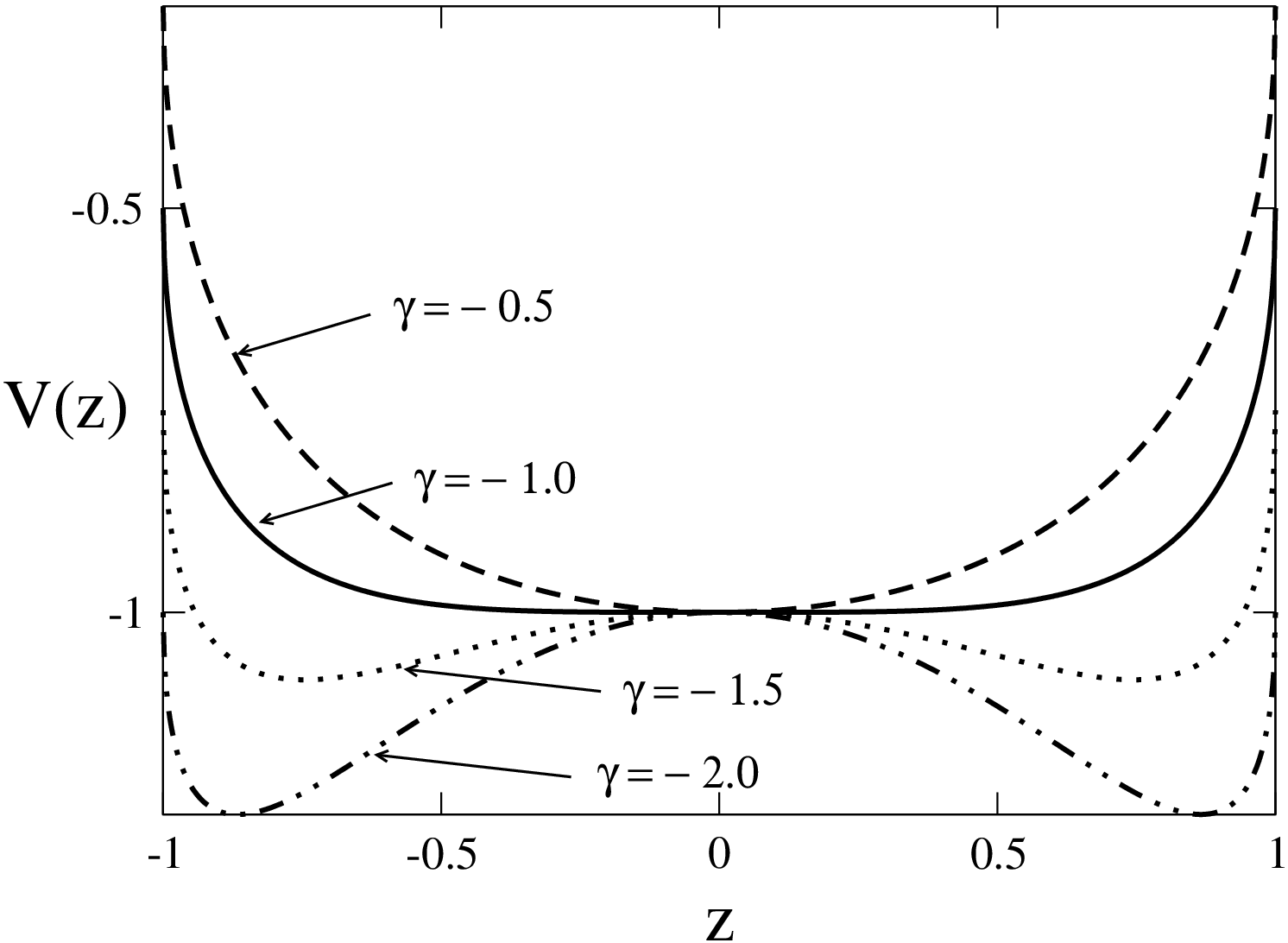}
\caption{Effective potential (\ref{veff}) for different values of $\gamma$.} \label{poten}
\end{figure}

In our opinion the real value of continuum approximation is that one can predict the properties of the system by simply analyzing a form of the effective potential. Figure~\ref{poten} presents the shape of the potential for three crucial regions: before ($\gamma>-1$), at ($\gamma=-1$), and beyond ($\gamma<-1$) the critical point. In the first region the potential has typically a quadratic form, while at the $\gamma=-1$ it broadens substantially and becomes quartic. In the last region it has a form of double well with the minima at $z=\pm z_0$. This analysis paints a picture of the second order quantum phase transition, with the critical point at $\gamma=-1$.

%%%%%%%%%%%%%%%%%%%%%%%%%%%%%%%%%%%%%%%%%%%%%%%%%%%%%%%%%%%%%%%%%%%%%%%%%%%%%%%%%%%%%%%%%
%\section{Fluctuations}
%\label{secfluct}
%In the previous sections we have introduced the $N$-conserving Bogoliubov description and
%the continuous approximation to the double-well problem. In the current section we will use these approaches to analyze properties of the system in a vicinity of the critical point.

The shape of the potential determines the shape of the ground state wave function, which is always symmetric. For $\gamma \geq -1$ it is bell shaped and centered around $z=0$ with the width depending on $1/N$. For $\gamma < -1$ the wave function has a double hump structure, centered around $\pm z_0$. This structure suggest the choice of an order parameter. Our choice is the absolute value of the population difference between the wells:
\be
\hat {\cal B}=\frac{1}{2}\left|\hat a_1^\dagger\hat a_1-\hat a_2^\dagger\hat a_2\right|,
\label{boper}
\ee
which in the continuum approximation is equal to $N|z|/2$.
In the classical theory the system would choose only one of the wells. Hence it would be natural to define $z$ as the order parameter. However in the quantum case the ground state of the system is the superposition of wavepackets localized in both wells. The modulus makes it insensitive to this superposition, it only examines the width of the wave-function.

\begin{figure}[tbp]
\includegraphics[width=9cm,angle=0]{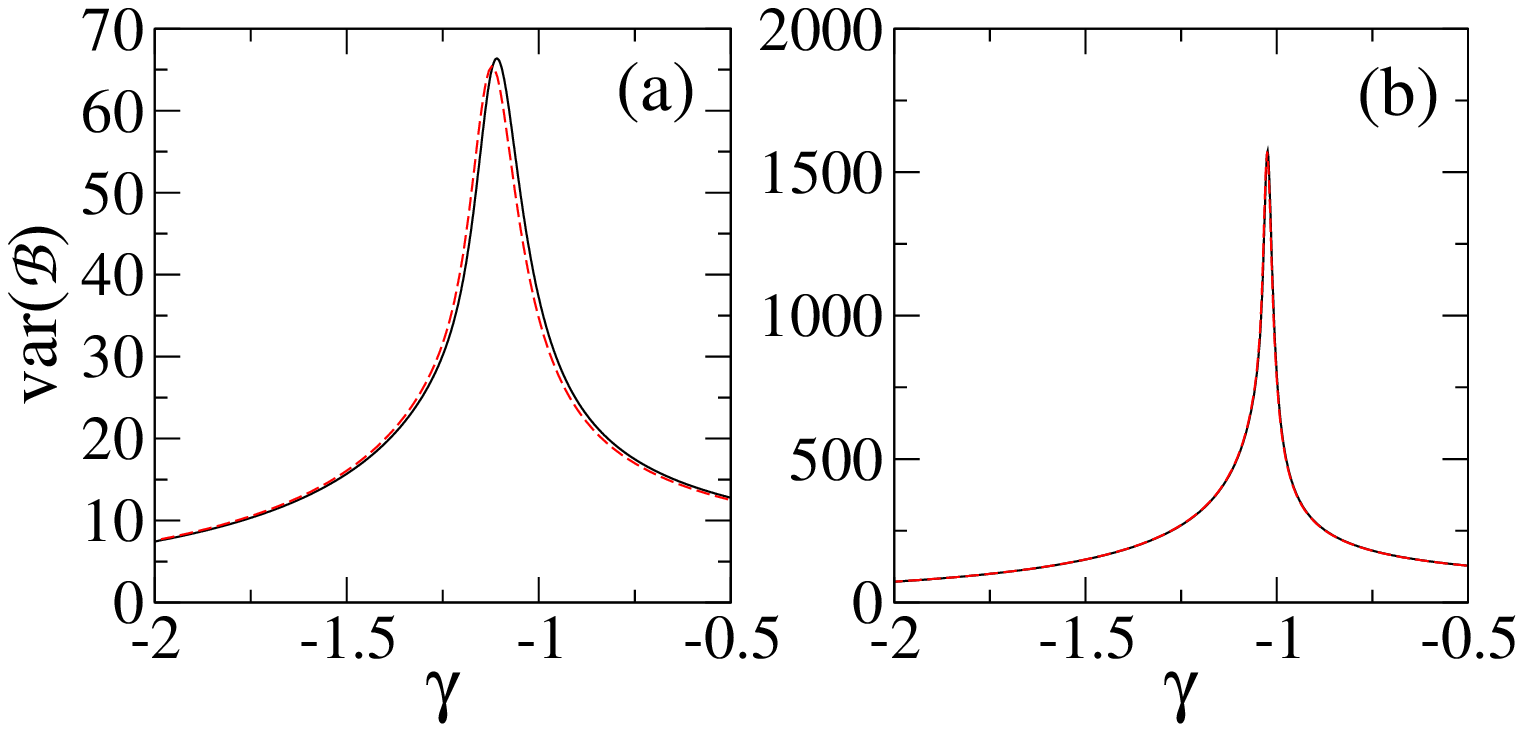}
\caption{(color online). Variance of the $\hat {\cal B}$ operator, Eq.(\ref{boper}), in the ground state of the double well problem obtained from exact diagonalization of the Hamiltonian (\ref{Ham2}) (solid black lines) and by solving Eq.(\ref{finaleq}) (dashed red lines) for $N=100$ (a) and $N=1000$ (b).}
\label{den2w}
\end{figure}

A clear sign of approaching the critical point is the increase of the fluctuations of the order parameter. The variance of $\hat {\cal B}$ in the ground state of the double well system versus $\gamma$ is shown in Fig.~\ref{den2w} for two different numbers of particles. Figure~\ref{den2w} indicates that the fluctuations are maximal around the critical point. With increasing value of $N$ the width of variance decreases and its maximum tends to $\gamma=-1$.

%Here we investigate fluctuation of number atoms in one of the wells. For $\gamma \rightarrow -\infty$ the ground state of the system approaches a Schr\"odinger cat state $1/\sqrt{2}(|0,N\rangle + |N,0\rangle )$, for which the fluctuations are the largest. This kind of fluctuations has been analyzed in details in Ref.~\cite{catstates}. To estimate the fluctuations where the {\it Schr\"odinger cat  effects} are eliminated one could consider slightly asymmetric double well problem. Alternatively one can choose another observable.  In the symmetric double well problem the ground state
%in the Fock basis $|N-n,n\ra$, Eq.~(\ref{fockb}), is symmetric, i.e. $\psi_n=\psi_{N-n}$,
%and for $\gamma<-1$ it reveals a double hump structure. The observable
%\be
%\hat {\cal B}=\frac{1}{2}\left|\hat a_1^\dagger\hat a_1-\hat a_2^\dagger\hat a_2\right|,
%\label{boper}
%\ee
%measures {\it the absolute value} of the difference of the population in both wells.
%
%By taking modulus we get rid of Schr\"odinger cat effect and is not sensitive to the sign. Thus, fluctuations
%of $\hat {\cal B}$ are the ones we are looking for. In the continuum approximation these fluctuations correspond to fluctuations of $N|z|/2$.

%%%%%%%%%%%%%%%%%%%%%%%%%%%%%%%%%%%%%%%%%%%%%%%%%%%%%%%%%%%%%%%%%%%%%%%%%%%
%\section{Conclusions}

In conclusion, we analyzed the behavior of the ground state of the Bose gas with attractive interactions in a symmetric double well potential. We showed that this system experiences second order quantum phase transition. This transition is revealed within the description using continuum approximation. This approximation allows for reduction of the many-body system in the two mode approximation (Bose-Hubbard model) to a problem of a fictitious quantum particle in an effective potential. Transformation of the effective potential from quadratic through quartic to double well with the increase of the interactions strength fits well to the Landau-Ginzburg scenario and allows to identify the order parameter and the critical point. The presence of the critical point manifests itself by sudden increase of the fluctuations of the order parameter (so called critical fluctuations), which we confirm numerically.

%%%%%%%%%%%%%%%%%%%%%%%%%%%%%%%%%%%%%%%%%%%%%%%%%%%%%%%%%%%%%%%%%%%%%%%%%%%
%\section*{ Acknowledgements }

The work of B. O. was supported by Polish Government scientific funds
(2008-2010) as a research project. K. S. was supported by Marie Curie ToK project
COCOS (MTKD-CT-2004-517186). P Z. and J. Ch. acknowledge the support of Polish Government scientific grant (2007-2010) and M. T. of Polish Government scientific grant (2006-2009).


\begin{thebibliography}{99}
\bibitem{remark1}
It is justified as along as $\psi_n$ is close to zero for $z_n\simeq\pm 1$. Numerical solutions of Eq.(\ref{recur}) show that this condition is satisfied except for the region $\gamma\ll-1$, when the strong condensate splitting ocours.
\end{thebibliography}
\end{document}